\documentclass[aip, apl,reprint]{revtex4-1}
\usepackage{graphicx}
\usepackage{dcolumn}
\usepackage{bm}
\usepackage{hyperref}
\usepackage{textcomp} 
\usepackage{gensymb} 
\usepackage{wasysym} 

\begin{document}

\title{All-optical lithography process for contacting atomically-precise devices}

\author{D. R. Ward}
\author{M. T. Marshall}
\author{D. M. Campbell}
\author{T. M. Lu}
\author{J. C. Koepke}
\author{D. A. Scrymgeour}
\author{E. Bussmann}
\author{S. Misra}
\email{smisra@sandia.gov}
\affiliation{Sandia National Laboratories, New Mexico 87185}

\date{\today}

\begin{abstract}
We describe an all-optical lithography process that can be used to make electrical contact to atomic-precision donor devices made in silicon using scanning tunneling microscopy (STM). This is accomplished by implementing a  cleaning procedure in the STM that allows the integration of metal alignment marks and ion-implanted contacts at the wafer level. Low-temperature transport measurements of a patterned device establish the viability of the process.  
\end{abstract}

\maketitle

The ability to fabricate devices with atomic precision holds promise for revealing the key physics underlying everything from quantum bits \cite{Buch,Watson} to ultra-scaled digital circuits \cite{Weber, Fueschle, Rudolph, Fuhrer, Levy}. A common atomic-precision fabrication (APFab) pathway uses a scanning tunneling microscope (STM) to create lithographic patterns on a hydrogen-passivated Si(100) surface \cite{Lyding}. Phosphine gas introduced into the vacuum system selectively adsorbs on sites where Si dangling bonds have been re-exposed by patterning \cite{Simmons}, yielding atomically precise, planar structures made of P donors. Unlike electron beam lithography (EBL), which can pattern hydrogen with a resolution of around 100~nm and is unable to image the pattern\cite{SimmonsEBL}, the STM is an ideal instrument for this process because it can both pattern and image the hydrogen resist with atomic precision \cite{Mechanism}. However, STMs are typically capable of patterning devices only up to 10~$\mu$m by 10~$\mu$m in size, which are too small to directly contact. A post-patterning microfabrication process, consisting of etching via holes in an encapsulating Si overlayer and then depositing  metal in direct contact with the planar donor layer, is used to make electrical contact to the devices. Even the largest features made with the STM are small enough that this contacting process relies on EBL for patterning and 200 nm- scale processing. At this scale, making good electrical contact between a deposited metal and an atomically-thin one-dimensional line of donors at the edge of an etched hole is challenging, and even successful EBL process flows in this application are rate-limiting. 

In this paper, we detail an all-optical lithography contacting process that reduces the time of fabricating an atomic-precision device by an order of magnitude. This is made possible by the integration of both ion-implanted contacts and metal alignment marks in the starting material, which bridge the scale between the largest regions accessible by STM and the smallest length scale accessible by low-cost photolithography. Specifically, the ion-implanted contacts neck down to a small enough area that the STM can place the APFab device in direct contact with them, and extend out to a region large enough that multiple photolithography steps done to a precision of 2~$\mu$m can connect the APFab device to 200~$\mu$m sized metal pads. Directly fabricating APFab donor structures on top of ion implanted Si simplifies the burden on microfabrication to that of  making contact between deposited metal and an ion implanted region, which can be done with near perfect yield. Moreover, this entire contacting process can be executed in a single day using tools available in most clean-rooms, and can be run on multiple chips in parallel. Moving forward, the increased throughput of our reliable all-optical contacting process promises to dramatically reduce the cost of making new discoveries using APFab devices. 

\begin{figure}
\includegraphics[width=3.36in]{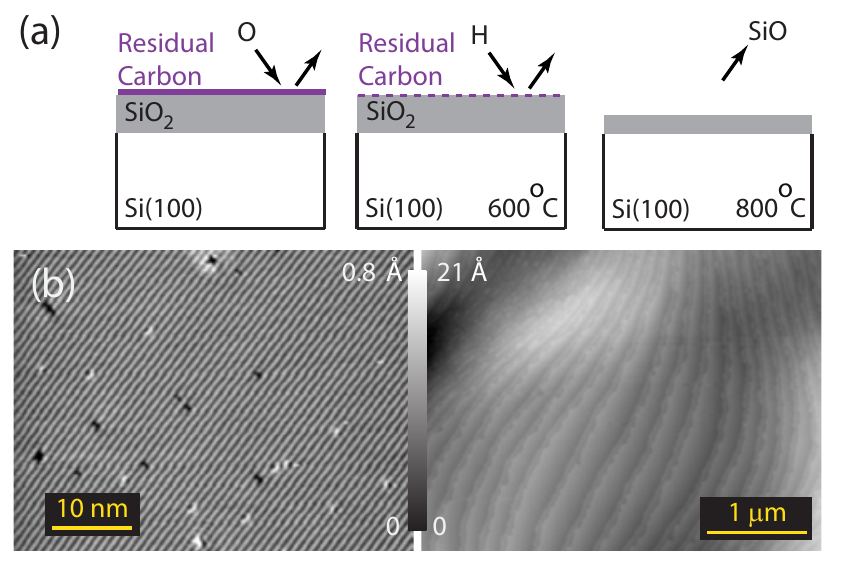}
\caption{(a) STM sample cleaning procedure, including oxygen plasma clean (left), atomic hydrogen clean (center), and oxide removal anneal (right). (b) STM topographs, taken at a tunnel junction setpoint of -2.5~V and 200~pA, show a clean Si(100) surface.}
\label{Fig1}
\end{figure}

The standard process for cleaning Si(100) samples for STM patterning  \cite{Brian} precludes the integration of metal alignment marks and ion implanted contacts, since the caustic chemical pre-cleaning {\it ex situ} and {\it in vacuo} annealing to 1200$\degree$C would destroy them. As illustrated in Figure \ref{Fig1}a, we have adopted a modified cleaning procedure and applied it to APFab. All the Si(100) samples used in this work were first cleaned {\it ex situ} in an ultrasonic bath of acetone and isopropyl alcohol to remove remnant photoresist. An oxygen plasma clean at 100~W of power for 20~minutes removes most of the remaining hydrocarbon debris from the surface. After inserting the samples into the STM vacuum chamber, we degas the samples by heating them successively to 450$\degree$C for 20  minutes and then 600$\degree$C for 40 minutes. Subsequently exposing the samples to atomic hydrogen removes the remaining trace carbon on the surface, based on the process described in Ref. \onlinecite{Hbomb}. For this, a tungsten filament at 1700$\degree$C in a background pressure of 5x10$^{-6}$~torr hydrogen gas generates atomic hydrogen while the sample is heated to 600$\degree$C for 20 minutes. Finally, we heat the sample to a modest 800$\degree$C to remove the surface oxide. The resultant surfaces are atomically clean on small length scales and show no contamination on larger length scales (Figure \ref{Fig1}b). 

\begin{figure}
\includegraphics[width=3.36in]{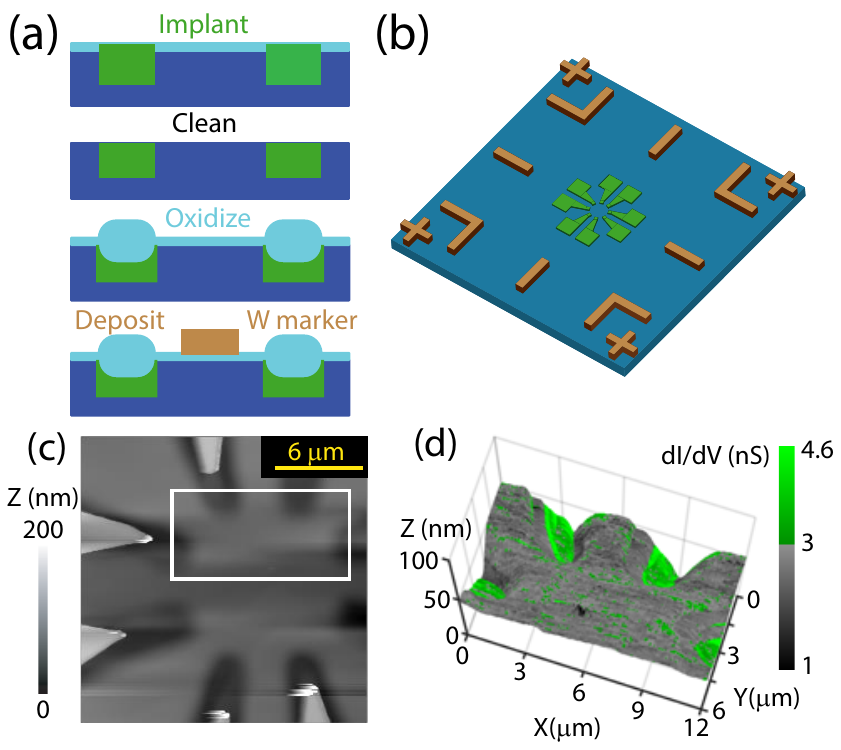}
\caption{(a) Side view of the process flow for integrating ion-implanted contacts and metal marks into the starting wafer. Here, Si is shown in blue, oxide in teal, implanted regions in green, and metal markers in brown. (b) Illustration of device fabrication process covering the 170 x 170 $\mu$m device area of the die, including the ion-implanted region (green) and the high contrast tungsten alignment markers (brown). (c) Atomic force microscope topographic image of the center of the implanted region of a chip after the sample preparation routine outlined in Figure \ref{Fig1}. STM data in (d) was taken in the white box before sample was removed from vacuum. The raised plateaus at the edge of the field of view are the remnant oxide from the middle of the ion implant. (d) False color overlay of STM differential conductance on simultaneously acquired topographic data of the ion-implanted region. Topographic data was taken with a tunnel junction setpoint of -1.5~V and 800~pA; the differential conductivity was recorded simultaneously using a lockin-amplifier at a frequency of 511~Hz and an ac excitation of 50~mV. The speckled green areas in between the implants are an artifact of noise- they do not appear in both the forward-scanned and backward-scanned frames of the data (not shown).}
\label{Fig2}
\end{figure}

This cleaning procedure enables the integration of ion-implanted contacts and metal alignment marks in the starting material, at the wafer level, as illustrated in Figure \ref{Fig2}a. Alignment marks are first etched in the material, a p-type Si(100) wafer having a volume resistivity  of 10-20~$\Omega$~cm and covered in 10~nm of sacrificial oxide (not shown). A photoresist mask is then used to perform a selective implant of As ions at 40~keV and an areal density of 3x10$^{15}$~ions/cm$^2$. These ion-implanted contacts start from a 40~$\mu$m by 40~$\mu$m field, large enough for aligning subsequent photolithography steps, and neck down to a 8~$\mu$m by 8~$\mu$m field, small enough for the STM to contact directly. The resist is then stripped and the sacrificial oxide removed using HF, exposing a pristine surface. Following RCA1, RCA2 , and HF cleans \cite{RCA}, a 10~nm steam oxide is grown at 850$\degree$C to protect the sample. Due to damage from the ion implantation, oxide grows at roughly 6 times the rate in implanted regions compared to pristine ones. The oxidation process is followed by a 15 minute anneal at 850$\degree$C in nitrogen. A final photoresist mask is used to deposit tungsten alignment markers (Figure \ref{Fig2}b). Critically for APFab, this process produces atomically clean surfaces once the resultant chips are subjected to the sample clean outlined above (Figure \ref{Fig1}b). By avoiding any acid-based cleaning, metal alignment marks are not damaged. Limiting the flash to 800$\degree$C prevents any significant diffusion of the ion-implanted contacts. While ion-implanted contacts have been implemented before \cite{TuckerImplant}, the process flow relied on interdigitated implanted contacts across large regions of the chip, and is thus limited to simple two-terminal devices. No such limitations exist for the process flow we have implemented.

\begin{figure*}
\includegraphics[width=6.68in]{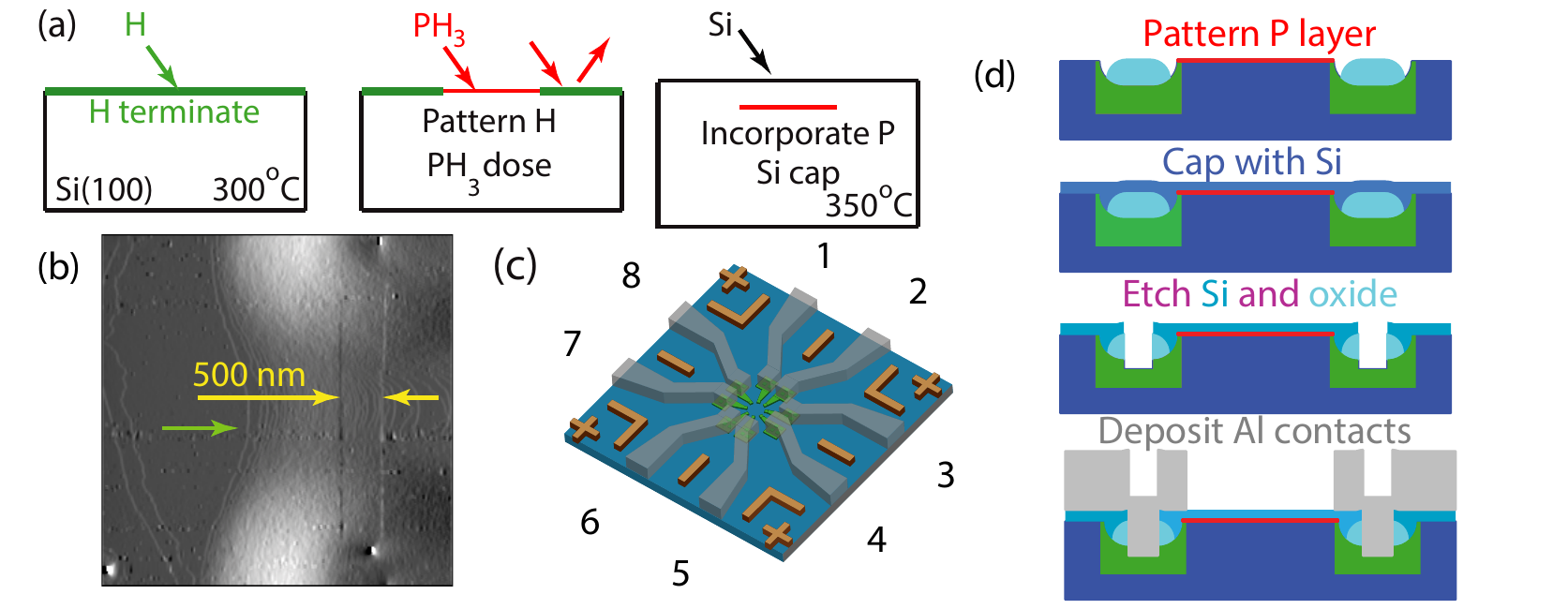}
\caption{(a) APFab process flow, including hydrogen termination (green), phosphine exposure (red) and silicon capping. (b) STM image, taken with a tunnel junction setpoint of -3~V and 200~pA, of a $\sim$2000~nm by 500~nm depassivated region. A Sobel filter has been used to enhance edge contrast. The green arrow points to a step edge, and yellow arrows point out the depassivated region connecting contacts 3 and 4 (right-side, top and bottom in c) (c and d) Plan and cross-sectional view of the process used to make contact to the APFab device. Here, Si is shown in blue, oxide in teal, implanted regions in green, metal markers in brown, the P APFab device in red, and aluminum contacts in grey. We wire bond directly to the aluminum contacts.}
\label{Fig3}
\end{figure*}

Flashing the sample to 800$\degree$C for 5 minutes is sufficient to remove the 10~nm of surface oxide, but leaves the implanted contacts buried in oxide. Since STM cannot tunnel into a thick insulator, the sample must be flashed for a longer period of time to expose enough doped Si to connect directly with hydrogen lithography. Figure \ref{Fig2}c shows a topographic image of a sample after it is flashed to 800$\degree$C for 15 minutes. Rather than a uniform reduction of all the thick oxide in the implanted region, the partial removal of  the thick oxide proceeds from the edge of the implanted region. This leaves a 30-60~nm deep trench that is easily identified by STM. Leveraging the high contrast metal markers and a high-resolution optical camera, we can align the tip precisely to the implanted contacts with sub 2~$\mu$m precision. To determine whether the implanted dopants diffuse out of the implanted region, we simultaneously acquired topographic and spectroscopic data using the STM in Figure \ref{Fig2}d. This data indicates that there is a region inside the trench that has an enhanced tunneling density of states, corresponding to a high concentration of activated donors \cite{LDOS}. Moreover, this enhanced density of states is sharply confined to the trench, indicating that As dopants have not diffused out from the implant region. 

Because the STM tip can be aligned to the high contrast metal marks using a long-focus optical microscope, hydrogen lithography can be used to make direct contact to the implanted contacts, which neck into a square that is 8~$\mu$m on a side. The samples are first terminated with atomic hydrogen, which serves as the monolayer resist for this process (Figure \ref{Fig3}a). The hydrogen termination is accomplished using a tungsten filament and filling the chamber to a pressure of 2x10$^{-6}$~torr, while holding the sample at 300$\degree$C. This hydrogen resist can both be imaged by STM at low junction biases (1-3~V), or removed at higher junction biases with either atomic precision ($\sim$3-5~V, 10~nm/s tip speed) or more coarsely ($\sim$7-10~V, 200~nm/s tip speed) \cite{Lyding, Mechanism, Zyvex}. In Figure \ref{Fig3}b, we have patterned a 500~nm wide wire between two implanted contacts. The resultant exposed dangling bonds selectively adsorb phosphine when it is introduced into the chamber \cite{Simmons}. We apply a total dose of 15~L at a chamber pressure of 2x10$^{-8}$~torr. A thermally activated surface decomposition reaction of the phosphine, at a temperature that leaves the hydrogen resist intact, results in P donors incorporated into the lattice at a density ranging between 17\% \cite{Simmons17} for the smallest windows (1 donor in a 3 dimer window) and 25\% for large areas \cite{Simmons25}. For the device in Figure \ref{Fig3}b, this will result in a P nanowire connecting the two As implanted regions. To preserve the atomically-precise donor-based device, it is then encapsulated in Si deposited at a rate of 1~nm/s to a thickness of 30~nm, while holding the sample at 350$\degree$C. From sample preparation to encapsulation, fabricating an atomic precision device that fans out to contact pads in a roughly 8~$\mu$m by 8~$\mu$m region takes about 12~hours, with the two most time-intensive parts being large STM scans to locate the implants, and patterning of the coarse features, which has been computer-automated \cite{SAND}.

The process of making electrical connection to the APFab device is now simplified as compared to EBL methods, requiring only optical lithography and standard clean-room microfabrication to put metal in direct contact with the eight ion-implanted contacts in a 40~$\mu$m by 40~$\mu$m area (Figure \ref{Fig3}d). Etching down to the implanted Si is complicated by the material stack in that part of the sample, which starts with the Si capping layer, followed by oxide which was incompletely removed during the sample flashing process, and finally by the doped Si. Contacts are made by patterning 2~$\mu$m diameter vias with optical lithography followed by a reactive ion etch of the Si capping layer using CF$_4$ at 25$\degree$C.  Next, the leads are patterned for a lift-off metal deposition.  Immediately before metal deposition a relatively long, 90~s etch in 1:6 BOE (buffered oxide etch) is used to remove the remaining oxide in the vias over the ion-implanted regions.  After depositing 150~nm of Al, by electron beam deposition, a standard metal lift-off process is used to complete the eight contacts that fan out into bond pads, shown schematically in  Figure \ref{Fig3}c. 

\begin{figure}
\includegraphics[width=3.36in]{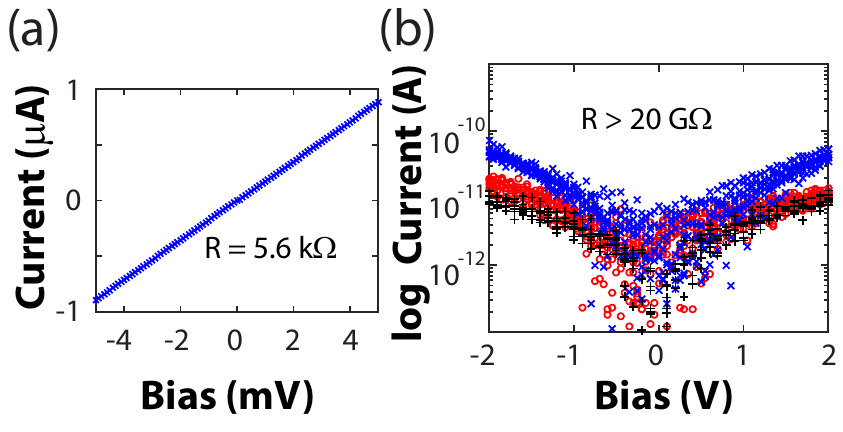}
\caption{(a) Current-voltage curve taken of the nanowire patterned in Figure \ref{Fig3}b between contacts 3 and 4. (b) Current-voltage curves taken between all other contact pairs on the same chip (blue), which had no P nanostructure connecting them. Also shown are current-voltage curves from a chip processed in the same way, but not dosed with phosphine (red), and a third chip that not subjected to any in-vacuum processing at all (black). All data were taken at 4~K by biasing a single contact and collecting the current on a neighboring contact while leaving the other contacts electrically floated. The absolute value of the current is plotted for negative biases.}
\label{Fig4}
\end{figure}

Electrical transport data on the simple nanowire device shown in Figure \ref{Fig3}b establishes the validity of this approach to making APFab devices (Figure \ref{Fig4}). These measurements must be carried out below $\sim$50~K to freeze out the carriers in the substrate.  At 4~K the only resistive elements in the path of the current are cables, the contacts to the 2D device layer, and the APFab nanowire itself. The DC transport through the nanowire is Ohmic down to tens of micro-Volts, and gives a resistance of 5.6~k$\Omega$. Accounting for the resistance of the nanowire (4 squares of P doped Si, whose resistivity is typically 525~$\Omega/\Box$), the total contact resistance to the P device layer comes to 1.75~k$\Omega$ per contact, which includes the metal-implant interface, the resistivity of the implanted region itself, and the implant-P interface.  Most importantly, we have found this method to produce a high yield of successful contacts; all eight contacts across three different chips have been shown to work in the same manner as Figure \ref{Fig4}. The high yield is attributed to the fact that implanted contacts represent an extended surface that deposited metal can come into contact with, including the bottom surface of the via holes. This is in contrast to trying to directly contact the two-dimensional P layer itself, in which case the metal needs to make a line contact at some location up the side-wall of the via hole. 

We also examine, in Figure \ref{Fig4}b, the transport between contacts that are not connected by a patterned APFab structure. These show a miniscule amount of leakage between isolated contact pairs- less than 0.1~nA at 2~V of bias. Two control samples- one which saw the same thermal processing as our APFab device but no phosphine dose, and a second one which was not subjected to any thermal processing- show similar levels of leakage current to one another. This indicates that the thermal budget of our process does not lead to enough As implant diffusion to be measurable. This also suggests that the additional leakage current between isolated contacts in the patterned sample originates largely from phosphine adsorbing through imperfections in the hydrogen resist. Both the Ohmic conduction through the nanowire, and the small leakage between unconnected pairs of contacts, compare well to an earlier effort which realized metal silicide contacts, but reported nonlinear I-V curves through an Ag nanowire with much larger leakage between unconnected contacts. \cite{Silicide} 

In conclusion, we have presented a new all-optical method for contacting APFab devices that takes a single day to execute, easily allows multiple chips to be processed in parallel, and achieves a high yield of successful contacts. This was made possible by adopting a process for cleaning Si(100) in an STM that is sufficiently low-temperature to allow for the integration of tungsten metal alignment markers and ion implanted contacts in the starting material.  The fast processing time and high yield are expected to dramatically reduce the costs of developing new innovations with APFab.

The Digital Electronics at the Atomic Limit project is supported by the Laboratory Directed Research and Development Program at Sandia National Laboratories, and was performed, in part, at the Center for Integrated Nanotechnologies, an Office of Science User Facility operated for the U.S. Department of Energy (DOE) Office of Science. Sandia National Laboratories is a multi-mission laboratory managed and operated by National Technology and Engineering Solutions of Sandia, LLC., a wholly owned subsidiary of Honeywell International, Inc., for the U.S. Department of Energy's National Nuclear Security Administration under contract DE-NA-0003525.

\bibliography{Manuscript}

\providecommand{\noopsort}[1]{}\providecommand{\singleletter}[1]{#1}%
\begin{thebibliography}{21}%
\makeatletter
\providecommand \@ifxundefined [1]{%
 \@ifx{#1\undefined}
}%
\providecommand \@ifnum [1]{%
 \ifnum #1\expandafter \@firstoftwo
 \else \expandafter \@secondoftwo
 \fi
}%
\providecommand \@ifx [1]{%
 \ifx #1\expandafter \@firstoftwo
 \else \expandafter \@secondoftwo
 \fi
}%
\providecommand \natexlab [1]{#1}%
\providecommand \enquote  [1]{``#1''}%
\providecommand \bibnamefont  [1]{#1}%
\providecommand \bibfnamefont [1]{#1}%
\providecommand \citenamefont [1]{#1}%
\providecommand \href@noop [0]{\@secondoftwo}%
\providecommand \href [0]{\begingroup \@sanitize@url \@href}%
\providecommand \@href[1]{\@@startlink{#1}\@@href}%
\providecommand \@@href[1]{\endgroup#1\@@endlink}%
\providecommand \@sanitize@url [0]{\catcode `\\12\catcode `\$12\catcode
  `\&12\catcode `\#12\catcode `\^12\catcode `\_12\catcode `\%12\relax}%
\providecommand \@@startlink[1]{}%
\providecommand \@@endlink[0]{}%
\providecommand \url  [0]{\begingroup\@sanitize@url \@url }%
\providecommand \@url [1]{\endgroup\@href {#1}{\urlprefix }}%
\providecommand \urlprefix  [0]{URL }%
\providecommand \Eprint [0]{\href }%
\providecommand \doibase [0]{http://dx.doi.org/}%
\providecommand \selectlanguage [0]{\@gobble}%
\providecommand \bibinfo  [0]{\@secondoftwo}%
\providecommand \bibfield  [0]{\@secondoftwo}%
\providecommand \translation [1]{[#1]}%
\providecommand \BibitemOpen [0]{}%
\providecommand \bibitemStop [0]{}%
\providecommand \bibitemNoStop [0]{.\EOS\space}%
\providecommand \EOS [0]{\spacefactor3000\relax}%
\providecommand \BibitemShut  [1]{\csname bibitem#1\endcsname}%
\let\auto@bib@innerbib\@empty
\bibitem [{\citenamefont {Buch}\ \emph {et~al.}(2013)\citenamefont {Buch},
  \citenamefont {Mahapatra}, \citenamefont {Rahman}, \citenamefont {Morello},\
  and\ \citenamefont {Simmons}}]{Buch}%
  \BibitemOpen
  \bibfield  {author} {\bibinfo {author} {\bibfnamefont {H.}~\bibnamefont
  {Buch}}, \bibinfo {author} {\bibfnamefont {S.}~\bibnamefont {Mahapatra}},
  \bibinfo {author} {\bibfnamefont {R.}~\bibnamefont {Rahman}}, \bibinfo
  {author} {\bibfnamefont {A.}~\bibnamefont {Morello}}, \ and\ \bibinfo
  {author} {\bibfnamefont {M.~Y.}\ \bibnamefont {Simmons}},\ }\href@noop {}
  {\bibfield  {journal} {\bibinfo  {journal} {Nature Comm.}\ }\textbf {\bibinfo
  {volume} {4}},\ \bibinfo {pages} {2017} (\bibinfo {year} {2013})}\BibitemShut
  {NoStop}%
\bibitem [{\citenamefont {Watson}\ \emph {et~al.}(2015)\citenamefont {Watson},
  \citenamefont {Weber}, \citenamefont {House}, \citenamefont {Buch},\ and\
  \citenamefont {Simmons}}]{Watson}%
  \BibitemOpen
  \bibfield  {author} {\bibinfo {author} {\bibfnamefont {T.~F.}\ \bibnamefont
  {Watson}}, \bibinfo {author} {\bibfnamefont {B.}~\bibnamefont {Weber}},
  \bibinfo {author} {\bibfnamefont {M.~G.}\ \bibnamefont {House}}, \bibinfo
  {author} {\bibfnamefont {H.}~\bibnamefont {Buch}}, \ and\ \bibinfo {author}
  {\bibfnamefont {M.~Y.}\ \bibnamefont {Simmons}},\ }\href@noop {} {\bibfield
  {journal} {\bibinfo  {journal} {Phys. Rev. Lett.}\ }\textbf {\bibinfo
  {volume} {115}},\ \bibinfo {pages} {166806} (\bibinfo {year}
  {2015})}\BibitemShut {NoStop}%
\bibitem [{\citenamefont {Weber}\ \emph {et~al.}(2012)\citenamefont {Weber},
  \citenamefont {Mahapatra}, \citenamefont {Ryu}, \citenamefont {an~A.~Fuhrer},
  \citenamefont {Reusch}, \citenamefont {Thompson}, \citenamefont {Lee},
  \citenamefont {Klimeck}, \citenamefont {Hollenberg},\ and\ \citenamefont
  {Simmons}}]{Weber}%
  \BibitemOpen
  \bibfield  {author} {\bibinfo {author} {\bibfnamefont {B.}~\bibnamefont
  {Weber}}, \bibinfo {author} {\bibfnamefont {S.}~\bibnamefont {Mahapatra}},
  \bibinfo {author} {\bibfnamefont {H.}~\bibnamefont {Ryu}}, \bibinfo {author}
  {\bibfnamefont {S.~L.}\ \bibnamefont {an~A.~Fuhrer}}, \bibinfo {author}
  {\bibfnamefont {T.~C.~G.}\ \bibnamefont {Reusch}}, \bibinfo {author}
  {\bibfnamefont {D.~L.}\ \bibnamefont {Thompson}}, \bibinfo {author}
  {\bibfnamefont {W.~C.~T.}\ \bibnamefont {Lee}}, \bibinfo {author}
  {\bibfnamefont {G.}~\bibnamefont {Klimeck}}, \bibinfo {author} {\bibfnamefont
  {L.~C.~L.}\ \bibnamefont {Hollenberg}}, \ and\ \bibinfo {author}
  {\bibfnamefont {M.~Y.}\ \bibnamefont {Simmons}},\ }\href@noop {} {\bibfield
  {journal} {\bibinfo  {journal} {Science}\ }\textbf {\bibinfo {volume}
  {335}},\ \bibinfo {pages} {64} (\bibinfo {year} {2012})}\BibitemShut
  {NoStop}%
\bibitem [{\citenamefont {Fueschle}\ \emph {et~al.}(2012)\citenamefont
  {Fueschle}, \citenamefont {Miwa}, \citenamefont {Mahapatra}, \citenamefont
  {Ryo}, \citenamefont {Lee}, \citenamefont {Warschkow}, \citenamefont
  {Hollenberg}, \citenamefont {Klimeck},\ and\ \citenamefont
  {Simmons}}]{Fueschle}%
  \BibitemOpen
  \bibfield  {author} {\bibinfo {author} {\bibfnamefont {M.}~\bibnamefont
  {Fueschle}}, \bibinfo {author} {\bibfnamefont {J.~A.}\ \bibnamefont {Miwa}},
  \bibinfo {author} {\bibfnamefont {S.}~\bibnamefont {Mahapatra}}, \bibinfo
  {author} {\bibfnamefont {H.}~\bibnamefont {Ryo}}, \bibinfo {author}
  {\bibfnamefont {S.}~\bibnamefont {Lee}}, \bibinfo {author} {\bibfnamefont
  {O.}~\bibnamefont {Warschkow}}, \bibinfo {author} {\bibfnamefont {L.~C.~L.}\
  \bibnamefont {Hollenberg}}, \bibinfo {author} {\bibfnamefont
  {G.}~\bibnamefont {Klimeck}}, \ and\ \bibinfo {author} {\bibfnamefont
  {M.~Y.}\ \bibnamefont {Simmons}},\ }\href@noop {} {\bibfield  {journal}
  {\bibinfo  {journal} {Nature Nano.}\ }\textbf {\bibinfo {volume} {7}},\
  \bibinfo {pages} {242} (\bibinfo {year} {2012})}\BibitemShut {NoStop}%
\bibitem [{\citenamefont {Rudolph}\ \emph {et~al.}(2014)\citenamefont
  {Rudolph}, \citenamefont {Carr}, \citenamefont {Subramania}, \citenamefont
  {Eyck}, \citenamefont {Dominguez}, \citenamefont {Pluym}, \citenamefont
  {Lilly}, \citenamefont {Carroll},\ and\ \citenamefont {Bussmann}}]{Rudolph}%
  \BibitemOpen
  \bibfield  {author} {\bibinfo {author} {\bibfnamefont {M.}~\bibnamefont
  {Rudolph}}, \bibinfo {author} {\bibfnamefont {S.~M.}\ \bibnamefont {Carr}},
  \bibinfo {author} {\bibfnamefont {G.}~\bibnamefont {Subramania}}, \bibinfo
  {author} {\bibfnamefont {G.~T.}\ \bibnamefont {Eyck}}, \bibinfo {author}
  {\bibfnamefont {J.}~\bibnamefont {Dominguez}}, \bibinfo {author}
  {\bibfnamefont {T.}~\bibnamefont {Pluym}}, \bibinfo {author} {\bibfnamefont
  {M.~P.}\ \bibnamefont {Lilly}}, \bibinfo {author} {\bibfnamefont {M.~S.}\
  \bibnamefont {Carroll}}, \ and\ \bibinfo {author} {\bibfnamefont
  {E.}~\bibnamefont {Bussmann}},\ }\href@noop {} {\bibfield  {journal}
  {\bibinfo  {journal} {Appl. Phys. Lett}\ }\textbf {\bibinfo {volume} {105}},\
  \bibinfo {pages} {163110} (\bibinfo {year} {2014})}\BibitemShut {NoStop}%
\bibitem [{\citenamefont {Pascher}\ \emph {et~al.}(2016)\citenamefont
  {Pascher}, \citenamefont {Hennel}, \citenamefont {Mueller},\ and\
  \citenamefont {Fuhrer}}]{Fuhrer}%
  \BibitemOpen
  \bibfield  {author} {\bibinfo {author} {\bibfnamefont {N.}~\bibnamefont
  {Pascher}}, \bibinfo {author} {\bibfnamefont {S.}~\bibnamefont {Hennel}},
  \bibinfo {author} {\bibfnamefont {S.}~\bibnamefont {Mueller}}, \ and\
  \bibinfo {author} {\bibfnamefont {A.}~\bibnamefont {Fuhrer}},\ }\href@noop {}
  {\bibfield  {journal} {\bibinfo  {journal} {New J. of Phys.}\ }\textbf
  {\bibinfo {volume} {18}},\ \bibinfo {pages} {083001} (\bibinfo {year}
  {2016})}\BibitemShut {NoStop}%
\bibitem [{\citenamefont {Cheng}\ \emph {et~al.}(2011)\citenamefont {Cheng},
  \citenamefont {Siles}, \citenamefont {Bi}, \citenamefont {Cen}, \citenamefont
  {Bogorin}, \citenamefont {Bark}, \citenamefont {Folkman}, \citenamefont
  {Park}, \citenamefont {Eom}, \citenamefont {Medeiros-Ribiero},\ and\
  \citenamefont {Levy}}]{Levy}%
  \BibitemOpen
  \bibfield  {author} {\bibinfo {author} {\bibfnamefont {G.}~\bibnamefont
  {Cheng}}, \bibinfo {author} {\bibfnamefont {P.~F.}\ \bibnamefont {Siles}},
  \bibinfo {author} {\bibfnamefont {F.}~\bibnamefont {Bi}}, \bibinfo {author}
  {\bibfnamefont {C.}~\bibnamefont {Cen}}, \bibinfo {author} {\bibfnamefont
  {D.~F.}\ \bibnamefont {Bogorin}}, \bibinfo {author} {\bibfnamefont {C.~W.}\
  \bibnamefont {Bark}}, \bibinfo {author} {\bibfnamefont {C.~M.}\ \bibnamefont
  {Folkman}}, \bibinfo {author} {\bibfnamefont {J.-W.}\ \bibnamefont {Park}},
  \bibinfo {author} {\bibfnamefont {C.-B.}\ \bibnamefont {Eom}}, \bibinfo
  {author} {\bibfnamefont {G.}~\bibnamefont {Medeiros-Ribiero}}, \ and\
  \bibinfo {author} {\bibfnamefont {J.}~\bibnamefont {Levy}},\ }\href@noop {}
  {\bibfield  {journal} {\bibinfo  {journal} {Nature Nano.}\ }\textbf {\bibinfo
  {volume} {6}},\ \bibinfo {pages} {343} (\bibinfo {year} {2011})}\BibitemShut
  {NoStop}%
\bibitem [{\citenamefont {Lyding}\ \emph {et~al.}(1994)\citenamefont {Lyding},
  \citenamefont {Shen}, \citenamefont {Hubacek}, \citenamefont {Tucker},\ and\
  \citenamefont {Abeln}}]{Lyding}%
  \BibitemOpen
  \bibfield  {author} {\bibinfo {author} {\bibfnamefont {J.~W.}\ \bibnamefont
  {Lyding}}, \bibinfo {author} {\bibfnamefont {T.-C.}\ \bibnamefont {Shen}},
  \bibinfo {author} {\bibfnamefont {J.~S.}\ \bibnamefont {Hubacek}}, \bibinfo
  {author} {\bibfnamefont {J.~R.}\ \bibnamefont {Tucker}}, \ and\ \bibinfo
  {author} {\bibfnamefont {G.~C.}\ \bibnamefont {Abeln}},\ }\href@noop {}
  {\bibfield  {journal} {\bibinfo  {journal} {Appl. Phys. Lett.}\ }\textbf
  {\bibinfo {volume} {64}},\ \bibinfo {pages} {2010} (\bibinfo {year}
  {1994})}\BibitemShut {NoStop}%
\bibitem [{\citenamefont {O'Brien}\ \emph {et~al.}(2001)\citenamefont
  {O'Brien}, \citenamefont {Chofield}, \citenamefont {Simmons}, \citenamefont
  {Clark}, \citenamefont {Dzurak}, \citenamefont {Curson}, \citenamefont
  {McAlpine}, \citenamefont {Hawley},\ and\ \citenamefont {Brown}}]{Simmons}%
  \BibitemOpen
  \bibfield  {author} {\bibinfo {author} {\bibfnamefont {J.~L.}\ \bibnamefont
  {O'Brien}}, \bibinfo {author} {\bibfnamefont {S.~R.}\ \bibnamefont
  {Chofield}}, \bibinfo {author} {\bibfnamefont {M.~Y.}\ \bibnamefont
  {Simmons}}, \bibinfo {author} {\bibfnamefont {R.~G.}\ \bibnamefont {Clark}},
  \bibinfo {author} {\bibfnamefont {A.~S.}\ \bibnamefont {Dzurak}}, \bibinfo
  {author} {\bibfnamefont {N.~J.}\ \bibnamefont {Curson}}, \bibinfo {author}
  {\bibfnamefont {N.~S.}\ \bibnamefont {McAlpine}}, \bibinfo {author}
  {\bibfnamefont {M.~E.}\ \bibnamefont {Hawley}}, \ and\ \bibinfo {author}
  {\bibfnamefont {G.~W.}\ \bibnamefont {Brown}},\ }\href@noop {} {\bibfield
  {journal} {\bibinfo  {journal} {Phys. Rev. B}\ }\textbf {\bibinfo {volume}
  {64}},\ \bibinfo {pages} {161401} (\bibinfo {year} {2001})}\BibitemShut
  {NoStop}%
\bibitem [{\citenamefont {Hallam}\ \emph {et~al.}(2007)\citenamefont {Hallam},
  \citenamefont {Butcher}, \citenamefont {Goh}, \citenamefont {Reuss},\ and\
  \citenamefont {Simmons}}]{SimmonsEBL}%
  \BibitemOpen
  \bibfield  {author} {\bibinfo {author} {\bibfnamefont {T.}~\bibnamefont
  {Hallam}}, \bibinfo {author} {\bibfnamefont {M.~J.}\ \bibnamefont {Butcher}},
  \bibinfo {author} {\bibfnamefont {K.~E.~J.}\ \bibnamefont {Goh}}, \bibinfo
  {author} {\bibfnamefont {F.~J.}\ \bibnamefont {Reuss}}, \ and\ \bibinfo
  {author} {\bibfnamefont {M.~Y.}\ \bibnamefont {Simmons}},\ }\href@noop {}
  {\bibfield  {journal} {\bibinfo  {journal} {J. Appl. Phys.}\ }\textbf
  {\bibinfo {volume} {102}},\ \bibinfo {pages} {034308} (\bibinfo {year}
  {2007})}\BibitemShut {NoStop}%
\bibitem [{\citenamefont {Shen}\ \emph {et~al.}(1995)\citenamefont {Shen},
  \citenamefont {Wang}, \citenamefont {Abeln}, \citenamefont {Tucker},
  \citenamefont {Lyding}, \citenamefont {Avouris},\ and\ \citenamefont
  {Walkup}}]{Mechanism}%
  \BibitemOpen
  \bibfield  {author} {\bibinfo {author} {\bibfnamefont {T.-C.}\ \bibnamefont
  {Shen}}, \bibinfo {author} {\bibfnamefont {C.}~\bibnamefont {Wang}}, \bibinfo
  {author} {\bibfnamefont {G.~C.}\ \bibnamefont {Abeln}}, \bibinfo {author}
  {\bibfnamefont {J.~R.}\ \bibnamefont {Tucker}}, \bibinfo {author}
  {\bibfnamefont {J.~W.}\ \bibnamefont {Lyding}}, \bibinfo {author}
  {\bibfnamefont {P.}~\bibnamefont {Avouris}}, \ and\ \bibinfo {author}
  {\bibfnamefont {R.~E.}\ \bibnamefont {Walkup}},\ }\href@noop {} {\bibfield
  {journal} {\bibinfo  {journal} {Science}\ }\textbf {\bibinfo {volume}
  {268}},\ \bibinfo {pages} {1590} (\bibinfo {year} {1995})}\BibitemShut
  {NoStop}%
\bibitem [{\citenamefont {B.~S.~Swartzentruber}\ and\ \citenamefont
  {Lagally}(1989)}]{Brian}%
  \BibitemOpen
  \bibfield  {author} {\bibinfo {author} {\bibfnamefont {M.~B.~W.}\
  \bibnamefont {B.~S.~Swartzentruber}, \bibfnamefont {Y.-W.~Mo}}\ and\ \bibinfo
  {author} {\bibfnamefont {M.~G.}\ \bibnamefont {Lagally}},\ }\href@noop {}
  {\bibfield  {journal} {\bibinfo  {journal} {J. Vac. Sci. Technol. A}\
  }\textbf {\bibinfo {volume} {7}},\ \bibinfo {pages} {2901} (\bibinfo {year}
  {1989})}\BibitemShut {NoStop}%
\bibitem [{\citenamefont {Assmuth}\ \emph {et~al.}(2007)\citenamefont
  {Assmuth}, \citenamefont {Stimpel-Lindner}, \citenamefont {Senftleben},
  \citenamefont {Bayerstadler}, \citenamefont {Siluma}, \citenamefont
  {Baumgartner},\ and\ \citenamefont {Eisele}}]{Hbomb}%
  \BibitemOpen
  \bibfield  {author} {\bibinfo {author} {\bibfnamefont {A.}~\bibnamefont
  {Assmuth}}, \bibinfo {author} {\bibfnamefont {T.}~\bibnamefont
  {Stimpel-Lindner}}, \bibinfo {author} {\bibfnamefont {O.}~\bibnamefont
  {Senftleben}}, \bibinfo {author} {\bibfnamefont {A.}~\bibnamefont
  {Bayerstadler}}, \bibinfo {author} {\bibfnamefont {T.}~\bibnamefont
  {Siluma}}, \bibinfo {author} {\bibfnamefont {H.}~\bibnamefont {Baumgartner}},
  \ and\ \bibinfo {author} {\bibfnamefont {I.}~\bibnamefont {Eisele}},\
  }\href@noop {} {\bibfield  {journal} {\bibinfo  {journal} {Appl. Surf. Sci.}\
  }\textbf {\bibinfo {volume} {253}},\ \bibinfo {pages} {8389} (\bibinfo {year}
  {2007})}\BibitemShut {NoStop}%
\bibitem [{\citenamefont {Kern}\ and\ \citenamefont {Pauotien}(1970)}]{RCA}%
  \BibitemOpen
  \bibfield  {author} {\bibinfo {author} {\bibfnamefont {W.}~\bibnamefont
  {Kern}}\ and\ \bibinfo {author} {\bibfnamefont {D.~A.}\ \bibnamefont
  {Pauotien}},\ }\href@noop {} {\bibfield  {journal} {\bibinfo  {journal} {RCA
  Rev.}\ }\textbf {\bibinfo {volume} {31}},\ \bibinfo {pages} {187} (\bibinfo
  {year} {1970})}\BibitemShut {NoStop}%
\bibitem [{\citenamefont {Kim}, \citenamefont {Kline},\ and\ \citenamefont
  {Tucker}(2005)}]{TuckerImplant}%
  \BibitemOpen
  \bibfield  {author} {\bibinfo {author} {\bibfnamefont {J.~C.}\ \bibnamefont
  {Kim}}, \bibinfo {author} {\bibfnamefont {J.~S.}\ \bibnamefont {Kline}}, \
  and\ \bibinfo {author} {\bibfnamefont {J.~R.}\ \bibnamefont {Tucker}},\
  }\href@noop {} {\bibfield  {journal} {\bibinfo  {journal} {Appl. Surf. Sci.}\
  }\textbf {\bibinfo {volume} {239}},\ \bibinfo {pages} {335} (\bibinfo {year}
  {2005})}\BibitemShut {NoStop}%
\bibitem [{\citenamefont {Studer}\ \emph {et~al.}(2013)\citenamefont {Studer},
  \citenamefont {Schofield}, \citenamefont {Hirjibehedin},\ and\ \citenamefont
  {Curson}}]{LDOS}%
  \BibitemOpen
  \bibfield  {author} {\bibinfo {author} {\bibfnamefont {P.}~\bibnamefont
  {Studer}}, \bibinfo {author} {\bibfnamefont {S.~R.}\ \bibnamefont
  {Schofield}}, \bibinfo {author} {\bibfnamefont {C.~F.}\ \bibnamefont
  {Hirjibehedin}}, \ and\ \bibinfo {author} {\bibfnamefont {N.~J.}\
  \bibnamefont {Curson}},\ }\href@noop {} {\bibfield  {journal} {\bibinfo
  {journal} {Appl. Phys. Lett.}\ }\textbf {\bibinfo {volume} {102}},\ \bibinfo
  {pages} {012107} (\bibinfo {year} {2013})}\BibitemShut {NoStop}%
\bibitem [{\citenamefont {Ballard}\ \emph {et~al.}(2013)\citenamefont
  {Ballard}, \citenamefont {Sisson}, \citenamefont {Owen}, \citenamefont
  {Owen}, \citenamefont {Fuchs}, \citenamefont {Randall},\ and\ \citenamefont
  {Ehr}}]{Zyvex}%
  \BibitemOpen
  \bibfield  {author} {\bibinfo {author} {\bibfnamefont {J.~B.}\ \bibnamefont
  {Ballard}}, \bibinfo {author} {\bibfnamefont {T.~W.}\ \bibnamefont {Sisson}},
  \bibinfo {author} {\bibfnamefont {J.~H.~G.}\ \bibnamefont {Owen}}, \bibinfo
  {author} {\bibfnamefont {W.~R.}\ \bibnamefont {Owen}}, \bibinfo {author}
  {\bibfnamefont {E.}~\bibnamefont {Fuchs}}, \bibinfo {author} {\bibfnamefont
  {J.~A. J.~N.}\ \bibnamefont {Randall}}, \ and\ \bibinfo {author}
  {\bibfnamefont {J.~R.~V.}\ \bibnamefont {Ehr}},\ }\href@noop {} {\bibfield
  {journal} {\bibinfo  {journal} {J. Vac. Sci. Tech. B}\ }\textbf {\bibinfo
  {volume} {31}},\ \bibinfo {pages} {06FC01} (\bibinfo {year}
  {2013})}\BibitemShut {NoStop}%
\bibitem [{\citenamefont {Schofield}\ \emph {et~al.}(2003)\citenamefont
  {Schofield}, \citenamefont {Curson}, \citenamefont {Simmons}, \citenamefont
  {Ruess}, \citenamefont {Hallam}, \citenamefont {Oberbeck},\ and\
  \citenamefont {Clark}}]{Simmons17}%
  \BibitemOpen
  \bibfield  {author} {\bibinfo {author} {\bibfnamefont {S.~R.}\ \bibnamefont
  {Schofield}}, \bibinfo {author} {\bibfnamefont {N.~J.}\ \bibnamefont
  {Curson}}, \bibinfo {author} {\bibfnamefont {M.~Y.}\ \bibnamefont {Simmons}},
  \bibinfo {author} {\bibfnamefont {F.~J.}\ \bibnamefont {Ruess}}, \bibinfo
  {author} {\bibfnamefont {T.}~\bibnamefont {Hallam}}, \bibinfo {author}
  {\bibfnamefont {L.}~\bibnamefont {Oberbeck}}, \ and\ \bibinfo {author}
  {\bibfnamefont {R.~G.}\ \bibnamefont {Clark}},\ }\href@noop {} {\bibfield
  {journal} {\bibinfo  {journal} {Phys. Rev. Lett.}\ }\textbf {\bibinfo
  {volume} {91}},\ \bibinfo {pages} {136104} (\bibinfo {year}
  {2003})}\BibitemShut {NoStop}%
\bibitem [{\citenamefont {Oberbeck}\ \emph {et~al.}(2002)\citenamefont
  {Oberbeck}, \citenamefont {Curson}, \citenamefont {Simmons}, \citenamefont
  {Brenner}, \citenamefont {Hamilton}, \citenamefont {Schofield},\ and\
  \citenamefont {Clark}}]{Simmons25}%
  \BibitemOpen
  \bibfield  {author} {\bibinfo {author} {\bibfnamefont {L.}~\bibnamefont
  {Oberbeck}}, \bibinfo {author} {\bibfnamefont {N.~J.}\ \bibnamefont
  {Curson}}, \bibinfo {author} {\bibfnamefont {M.~Y.}\ \bibnamefont {Simmons}},
  \bibinfo {author} {\bibfnamefont {R.}~\bibnamefont {Brenner}}, \bibinfo
  {author} {\bibfnamefont {A.~R.}\ \bibnamefont {Hamilton}}, \bibinfo {author}
  {\bibfnamefont {S.~R.}\ \bibnamefont {Schofield}}, \ and\ \bibinfo {author}
  {\bibfnamefont {R.~G.}\ \bibnamefont {Clark}},\ }\href@noop {} {\bibfield
  {journal} {\bibinfo  {journal} {Appl. Phys. Lett.}\ }\textbf {\bibinfo
  {volume} {81}},\ \bibinfo {pages} {3197} (\bibinfo {year}
  {2002})}\BibitemShut {NoStop}%
\bibitem [{\citenamefont {Ward}\ \emph {et~al.}(2016)\citenamefont {Ward},
  \citenamefont {Misra}, \citenamefont {Scrymgeour}, \citenamefont {Simonson},
  \citenamefont {Marshall}, \citenamefont {Koepke}, \citenamefont {Bussmann},
  \citenamefont {Carroll}, \citenamefont {Ballard}, \citenamefont {Owen},
  \citenamefont {Schmucker}, \citenamefont {Fuchs}, \citenamefont {Pryadkin},\
  and\ \citenamefont {Randall}}]{SAND}%
  \BibitemOpen
  \bibfield  {author} {\bibinfo {author} {\bibfnamefont {D.~R.}\ \bibnamefont
  {Ward}}, \bibinfo {author} {\bibfnamefont {S.}~\bibnamefont {Misra}},
  \bibinfo {author} {\bibfnamefont {D.~A.}\ \bibnamefont {Scrymgeour}},
  \bibinfo {author} {\bibfnamefont {R.~J.}\ \bibnamefont {Simonson}}, \bibinfo
  {author} {\bibfnamefont {M.~T.}\ \bibnamefont {Marshall}}, \bibinfo {author}
  {\bibfnamefont {J.~C.}\ \bibnamefont {Koepke}}, \bibinfo {author}
  {\bibfnamefont {E.}~\bibnamefont {Bussmann}}, \bibinfo {author}
  {\bibfnamefont {M.~S.}\ \bibnamefont {Carroll}}, \bibinfo {author}
  {\bibfnamefont {J.~B.}\ \bibnamefont {Ballard}}, \bibinfo {author}
  {\bibfnamefont {J.~H.~G.}\ \bibnamefont {Owen}}, \bibinfo {author}
  {\bibfnamefont {S.~W.}\ \bibnamefont {Schmucker}}, \bibinfo {author}
  {\bibfnamefont {E.}~\bibnamefont {Fuchs}}, \bibinfo {author} {\bibfnamefont
  {S.}~\bibnamefont {Pryadkin}}, \ and\ \bibinfo {author} {\bibfnamefont
  {J.~N.}\ \bibnamefont {Randall}},\ }\href@noop {} {\bibfield  {journal}
  {\bibinfo  {journal} {Sandia Tech. Rep.}\ }\textbf {\bibinfo {volume}
  {SAND2016}},\ \bibinfo {pages} {5470 C} (\bibinfo {year} {2016})}\BibitemShut
  {NoStop}%
\bibitem [{\citenamefont {Fujimori}\ \emph {et~al.}(2004)\citenamefont
  {Fujimori}, \citenamefont {Heike}, \citenamefont {Terada},\ and\
  \citenamefont {Hashizume}}]{Silicide}%
  \BibitemOpen
  \bibfield  {author} {\bibinfo {author} {\bibfnamefont {M.}~\bibnamefont
  {Fujimori}}, \bibinfo {author} {\bibfnamefont {S.}~\bibnamefont {Heike}},
  \bibinfo {author} {\bibfnamefont {Y.}~\bibnamefont {Terada}}, \ and\ \bibinfo
  {author} {\bibfnamefont {T.}~\bibnamefont {Hashizume}},\ }\href@noop {}
  {\bibfield  {journal} {\bibinfo  {journal} {Nanotech.}\ }\textbf {\bibinfo
  {volume} {15}},\ \bibinfo {pages} {S333} (\bibinfo {year}
  {2004})}\BibitemShut {NoStop}%
\end{thebibliography}%

\end{document}